\documentclass[useAMS,usedcolumn,usegraphicx,usenatbib]{mn2e}

\usepackage{times}
\usepackage{amsmath}                  % mathematischer Zeichensatz
\usepackage{amssymb}                  % mathematische Symbole
\usepackage{graphicx}
\usepackage{units}
\usepackage{dcolumn}
\usepackage{longtable}
\usepackage{lscape}
\usepackage{booktabs}
\usepackage{subfigure}
\usepackage{natbib}
\usepackage{aas_macros}
\usepackage[draft,pdftex,pdfpagemode={UseOutlines},bookmarks,bookmarksopen,colorlinks,linkcolor={blue},citecolor={green},urlcolor={red}]{hyperref}
\makeatletter
\newcolumntype{d}[1]{>{\DC@{.}{{.}}{#1}}c<{\DC@end}}
\newcolumntype{o}[1]{>{\DC@{+}{\pm}{#1}}c<{\DC@end}}
\makeatother
\makeatletter
\newcolumntype{f}[1]{>{\DC@{p}{\ldots}{#1}}c<{\DC@end}}
\makeatother

\let\orgautoref\autoref

\renewcommand{\autoref}
        {\def\equationautorefname{equation}%
         \def\figureautorefname{Fig.}%
         \def\subfigureautorefname{Fig.}%
         \def\partautorefname{part}%
         \def\chapterautorefname{chapter}%
         \def\sectionautorefname{section}%
         \def\subsectionautorefname{section}%
         \def\subsubsectionautorefname{section}%
         \def\appendixautorefname{appendix}%
         \def\Itemautorefname{item}%
         \def\tableautorefname{Table}%
         \def\lstlistingautorefname{Listing}%
         \orgautoref}
\providecommand{\autorefs}
        {\def\equationautorefname{equations}%
         \def\figureautorefname{Figs.}%
         \def\subfigureautorefname{Figs.}%
         \def\partautorefname{parts}%
         \def\chapterautorefname{chapters}%
         \def\sectionautorefname{sections}%
         \def\subsectionautorefname{sections}%
         \def\subsubsectionautorefname{sections}%
         \def\Itemautorefname{items}%
         \def\tableautorefname{Tables}%
         \def\lstlistingautorefname{Listings}%
         \orgautoref}

\newcommand{\rxja}{RX\,J1856.5-3754}
\newcommand{\rxjn}{RX\,J0720.4-3125}

\title[The origin of \rxja{} and \rxjn{} -- updated]{The origin of \rxja{} and \rxjn{} -- updated using new parallax measurements}
\author[N. Tetzlaff et al.]{N. Tetzlaff$^{1}$\thanks{E-mail:
nina@astro.uni-jena.de}, T. Eisenbeiss$^{1}$, R. Neuh\"auser$^{1}$ and M. M. Hohle$^{1,2}$\\
$^{1}$Astrophysikalisches Institut und Universit\"ats-Sternwarte Jena, Schillerg\"asschen 2-3, 07745 Jena, Germany\\
$^{2}$Max-Planck-Institut f\"ur extraterrestrische Physik, Giessenbachstra{\ss}e, 85741 Garching, Germany}

\begin{document}

\date{Accepted. Received; in original form}

\pagerange{\pageref{firstpage}--\pageref{lastpage}} \pubyear{2011}

\maketitle

\label{firstpage}

\begin{abstract} 
\rxja{} and \rxjn{} are the only young isolated radio-quiet neutron stars (NSs) for which trigonometric parallaxes were measured. Due to detection of their thermal emission in X-rays they are important to study NS cooling and to probe theoretical cooling models. Hence, a precise determination of their age is essential.\\
Recently, new parallax measurements of \rxja{} and \rxjn{} were obtained. Considering that NSs may originate from binary systems that got disrupted due to an asymmetric supernova, we attempt to identify runaway stars which may have been former companions to the NS progenitors. Such an identification would strongly support a particular birth scenario with time and place. \\
We trace back each NS, runaway star and the centres of possible birth associations (assuming that most NSs are ejected directly from their parent association) to find close encounters. The kinematic age is then given by the time since the encounter. We use Monte Carlo simulations to account for observational uncertainties and evaluate the outcome statistically.\\
Using the most recent parallax measurement of $\unit[8{.}16\pm0{.}80]{mas}$ for \rxja{} by \citeauthor{2010ApJ...724..669W}, we find that it originated in the Upper Scorpius association $\unit[0{.}46\pm0{.}05]{Myr}$ ago. This kinematic age is slightly larger than the value we reported earlier ($\unit[0{.}3]{Myr}$) using the old parallax value of $\unit[5{.}6\pm0{.}6]{mas}$ by \citeauthor{2003conf...Kaplan}. Our result is strongly supported by its current radial velocity that we predict to be $\unit[6^{+19}_{-20}]{km/s}$. This implies an inclination angle to the line-of sight of $\unit[88\pm6]{deg}$ consistent with estimates by \citeauthor{2001A&A...380..221V} from the bow shock. No suitable runaway star was found to be a potential former companion of \rxja{}.\\
Making use of a recent parallax measurement for \rxjn{} of $\unit[3{.}6\pm1{.}6]{mas}$ by \citeauthor{ThomasPhD}, we find that this NS was possibly born in Trumpler 10 $\unit[0{.}85\pm0{.}15]{Myr}$ ago. This kinematic age is somewhat larger than the one obtained using the old parallax value of $\unit[2{.}77\pm1{.}29]{mas}$ by \citeauthor{2007ApJ...660.1428K} ($\unit[0{.}5]{Myr}$). We suggest the B0 runaway supergiant HIP 43158 as a candidate for a former companion of the progenitor star. Then, the current distance of \rxjn{} to the Sun should be $\unit[286^{+27}_{-23}]{pc}$, in agreement with recent measurements. We then expect the radial velocity of \rxjn{} to be $\unit[-76^{+34}_{-17}]{km/s}$.
\end{abstract}


%%%%%%%%%%%%%%%%%%%%%%%%%%%%%%%%%%%%%%%%%%%%%%%%%%%%%%%%%%%%%%%%%%%%%%%%%%%%%%%%%%%%%%%%%%%%%%%%%%%%%%%%%%%%%%%%%%%%%%%%%%%%%%%%%%%%%%%%%%%%%%%%%

\section{Introduction}\label{sec:intro}

Neutron stars (NSs) show large proper motions which, with known distances, indicate high space velocities \citep[e.g.][]{1994Natur.369..127L,1997MNRAS.289..592L,1997MNRAS.291..569H,1998ApJ...505..315C,2002ApJ...568..289A,2005MNRAS.360..974H}. Some NSs even show velocities of the order of $\approx\unit[1000]{km/s}$ (e.g. PSR B1508+55, \citealt{2005ApJ...630L..61C}; PSR B2223+65, \citealt{1993MNRAS.261..113H}; \citealt{1993ApJ...411..674T}; RX J0822-4300, \citealt{2006A&A...457L..33H}, \citealt{2007ApJ...670..635W}). Those high velocities may be the result of asymmetric supernova (SN) explosions assigning the new-born NS a kick velocity for that a number of mechanisms have been suggested \citep[e.g.][]{1996PhRvL..76..352B,1996A&A...306..167J,2005ASPC..332..363J,2006ApJ...639.1007W,2009arXiv0906.2802K}. Another possibility is that the high-velocity NSs are the remnants of (symmetric\footnote{Although numerous three-dimensional simulations showed that SNe are most likely asymmetric \citep[e.g.][]{2010ApJ...725L.106W,2010ApJ...714.1371H,2008PhRvD..78f4056D,2008AIPC..983..369J,2006A&A...457..963S,2007Natur.445...58B}.}) SN explosions of the so-called hyper-velocity runaway stars
\citep{2007A&A...470L...9G,2008MNRAS.385..929G,2009MNRAS.395L..85G} which were ejected due to dynamical three- or four-body encounters either from the Galactic Centre \citep{1988Natur.331..687H} or from massive star clusters in the Galactic disk.\\
About $\unit[30]{\%}$ of young stars show different velocity properties than normal Population I stars \citep[][thereafter T11, a catalogue of young runaway stars]{1991AJ....102..333S,1993ASPC...35..207B,2011MNRAS.410..190T}. Two scenarios are accepted to produce those so-called ``runaway stars'' \citep{1961BAN....15..265B}. The binary-SN scenario \citep{1961BAN....15..265B} is related to the formation of the high velocity NSs (but note that high-velocity NSs may also be the result of an asymmetric SN explosion of a single massive star or a SN of a massive runaway star, see above): The runaway and NS are the products of a SN within a binary system. The velocity of the former secondary may be as large as its original orbital velocity \citep{1998A&A...330.1047T}. Runaway stars produced in this scenario should share typical properties such as a high rotational velocity $v\sin i$ and an enhanced helium abundance owing to momentum and mass transfer during binary evolution \citep{1993ASPC...35..207B}. We refer to such runaway stars as BSS runaway stars. The second scenario is the dynamical ejection due to gravitational interactions between massive stars in dense clusters \citep{1967BOTT....4...86P}.\\
Several studies have been made to investigate the origin of runaway stars \citep[e.g.][]{1961BAN....15..265B,1986ApJS...61..419G,2001A&A...365...49H,2005A&A...437..247D,2008A&A...489..105S,2008A&A...490.1071G} but only a few regarding the origin of fast-moving NSs \citep[e.g.][thereafter T10]{2001A&A...365...49H,2004ApJ...610..402V,2010MNRAS.402.2369T}.\\
In this paper we will re-investigate the origin of two isolated radio-quiet X-ray emitting NSs, namely \rxja{} and \rxjn{}. So far, only seven such sources have been identified for which they were named ``The Magnificent Seven'' (M7) (\citealt{2001ASPC..234..225T}; for recent reviews see \citealt{2007Ap&SS.308..181H,2008AIPC..968..129K}). They are bright X-ray sources associated with faint blue optical counterparts arising from a hot cooling surface. In only the two cases studied in this paper a trigonometric parallax was obtained. The M7, and especially \rxja{} and \rxjn{}, are very important NSs since brightness, parallax and temperature yield the size of the emitting area and hence, their (model dependent) radii. From spectra, one can in principle determine their masses and atmosphere composition, which eventually may lead to constraints on the equation of state.  With known luminosity and age, cooling curves can be verified. As the characteristic age represents only a rough estimate of the true age \citep[e.g.][]{1988MNRAS.234P..57B,2000Natur.406..158G,2002ApJ...567L.141M}, the kinematic age is very important to have a better estimate of the true age. Also, the characteristic age is significantly influenced by pulsar winds \citep[e.g.][]{2003A&A...409..641W} and possibly by emission of gravitational waves \citep[e.g.][]{2008CQGra..25w5011W}.\\
Under the assumption that most NSs form in associations or clusters of massive stars, the identification of the birth association of a NS by its flight path is possible if the uncertainties in the distance are moderate. Our assumption is justified since we observe associations and clusters of massive stars (also, the dispersion time scale of a massive star cluster is much longer than the lifetime of its most massive member stars), i.e. only a small fraction of their massive member stars is ejected from their parent cluster due to gravitational interactions before they end their lives in SNe (about 20-30 per cent of O and early B type stars are located outside of clusters, \citealt[e.g.][]{1998AJ....115..821M,2004ApJS..151..103M}, i.e. $\gtrsim$70 per cent of a cluster's member stars remain in the cluster). There are certainly NSs that form outside their parent cluster as there are massive runaway stars that will later explode in a SN event and become NSs. For the latter, the identification of their parent associations or birth sites is hardly possible. Also, it is possible, that a massive binary was ejected from its parent cluster. If the primary experiences a SN, the secondary might be ejected from the system and will later explode in a SN. For NSs whose progenitors experienced this so-called two-step-ejection scenario \citep{2010MNRAS.404.1564P}, it is not possible to identify their formation sites. However, this scenario applies only to 1-4 per cent of O stars \citep{2010MNRAS.404.1564P}. NSs formed from runaway stars may show a higher space velocity than those formed in their parent cluster since their velocity vector is a superposition of the runaway star's velocity and the kick velocity the NS receives at birth.\\
When searching for the parent association of a NS (under the above assumptions), the result is often not unique (T10). For that reason, it is desirable to find a second indicator for the location and time of the past SN, e.g. a possible former companion star that is now a runaway star.\\
Here, we note that not for every NS a (identifiable) runaway star must exist. The NS progenitor could have been a single star (possibly a runaway star, see above); a former massive companion could have already undergone a SN and is now a NS; or the former companion that is now a runaway star was not yet identified as such because, e.g., its absolute velocity is low \citep[e.g.][]{1998A&A...330.1047T} or the direction of its velocity vector is not significantly different from those of its neighbouring stars.\\

After investigating possible parent associations for \rxja{} and \rxjn{} in \autorefs{subsec:1856parentassoc} and \ref{subsec:0720parentassoc}, respectively, applying the procedure described in \autoref{sec:procedure} and using most recent parallax measurements, we attempt to identify the possible former companion for \rxja{} and \rxjn{} in \autorefs{subsec:1856comp} and \ref{subsec:0720comp}, respectively. We summarise our results and draw our conclusions in \autoref{sec:summary}.

%__________________________________________________________________

\section{Procedure}\label{sec:procedure}

Basically, we applied the same procedure as already carried out in T10, to that we refer for details (details may also be found in the additional online supporting information\footnote{http://www.astro.uni-jena.de/$\sim$nina/supporting\_info.pdf}). We perform Monte Carlo simulations to account for the uncertainties of the observables and the unknown radial velocity of NSs for which we assume a probability distribution derived from the one of pulsar space velocities by \citet{2005MNRAS.360..974H}.\\

In this paper, we extended the sample of 140 young associations and clusters given in T10 to a sample of 295 young associations and clusters within $\unit[3]{kpc}$ from the Sun. This sample of young associations and clusters contains 140 associations and clusters listed in T10, four additional young nearby associations from \citet{2008hsf2.book..757T} (Columba, Carina, Octans, Argus), 101 clusters with at least one star with a spectral type earlier B3 in the WEBDA database\footnote{http://www.univie.ac.at/webda/webda.html; \citet{2003A&A...410..511M}.} as well as 50 additional young clusters with kinematic data available in cluster catalogues (WEBDA; \citealt{2005A&A...440..403K,2007AN....328..889K,2010yCat....102022D}). \\

The distribution of separations $d_{min}$ (that is the smallest separation between the NS and the association/ cluster centre found in each Monte Carlo run) is supposed to obey a distribution of absolute differences of two 3D Gaussians (equation 1 in T10; equation A3 in \citealt{2001A&A...365...49H}; see also additional online supporting material). For investigating encounters with runaway stars, we calculate $d_{min}$ between the NS and the runaway star. If the two stars were at the same time and place, the $d_{min}$ distribution will follow equation 2 in T10 (equation A4 in \citealt{2001A&A...365...49H}).\\
We will adapt the theoretical formulae only to the first part of the $d_{min}$ distribution such that the slope and peak can be explained. The parameter $\mu$ then gives the positional difference between the two objects. The error of this difference, $\Delta$, can be estimated as $\Delta^2=\sigma_1^2+\sigma_2^2=2\sigma^2$.\\

To associate the NS/runaway star encounter position with an association/cluster, the trajectory of the latter is calculated simultaneously. Runaway star data are taken from T11 (and references therein, mainly \citealt{2007AA...474..653V}) for 2547 runaway stars (1705\footnote{In T11, 1703 stars are listed with radial velocity ($v_r$) measurements. For two further stars, we find $v_r$ in the literature: HIP 17158 with $v_r=\unit[55{.}19\pm0{.}73]{km/s}$ \citep{1995AN....316..209J} and HIP 68281 with $v_r=\unit[-11]{km/s}$ \citep{1983AJ.....88.1349S}.} with full 3D kinematics).\\
All calculations of trajectories are performed in a coordinate system centred on the Sun at present. We account for Solar motion using a local standard of rest of $\left(UVW\right)_\odot=\left(10{.}4\pm0{.}4,11{.}6\pm0{.}2,6{.}1\pm0{.}2\right)\,\mathrm{km/s}$ (T11), where $U$, $V$, $W$ are the velocity components in right handed Cartesian coordinates.\\

In general, we first perform $10^4$ Monte Carlo runs for each NS/association pair and $10^3$ Monte Carlo runs for each NS/runaway star pair (for the latter less runs are still sufficient due to the smaller errors on the runaway star kinematics compared to the dispersion of the association velocities) to find those associations and runaway stars that potentially crossed the past path of the NS, i.e. those for which the smallest $d_{min}$ value found in the calculations is less than three times the association radius (for NS/association pairs) or less than $\unit[10]{pc}$ (for NS/runaway star pairs), respectively. The latter value is justified since even smaller $d_{min}$ values are expected to be found after $10^3$ runs if the NS and the runaway star once were at the same place (see additional online supporting information). Those associations and runaway stars that fulfilled those conditions are then selected for a more detailed investigation (one to three million Monte Carlo runs). The outcome of these simulations is then discussed in detail. Regarding associations, we search for those for which the NS could have been within the association boundaries in the past while for runaway stars, we are looking for those runaway stars for which the NS and the runaway star might have been at the same place in the past, hence the distribution (slope) should obey equation 2 in T10. After three million runs the smallest $d_{min}$ value found is expected to be smaller than one parsec (see additional online supporting information). If this criterion is satisfied, we adapt the theoretically expected distributions to the first part of the ``observed'' $d_{min}$ distribution to explain its slope and peak. If we find $\mu\approx0$, we select the runaway star to be a former companion candidate. 

\section{Results}

\subsection{Identifying the parent association of \rxja{}}\label{subsec:1856parentassoc}

\citet{2010ApJ...724..669W} recently reported a parallactic distance of \rxja{} of $\unit[123^{+11}_{-15}]{pc}$ confirming earlier measurements by \citet{2002ApJ...576L.145W}. This distance is significantly smaller than the value of $\unit[178\pm20]{pc}$ claimed by \citet{2003conf...Kaplan} and used by T10 to evaluate the birth place of \rxja{}. Hence, it is worthwhile to re-investigate the origin of \rxja{}.\\
We adopt the following parameters for \rxja{} for the right ascension $\alpha$, declination $\delta$, parallax $\pi$ and proper motion $\mu$ \citep{2010ApJ...724..669W}:

\begin{equation}
	\begin{array}{l c l}
	\alpha &=& 18^\mathrm{h}56^\mathrm{m}35^\mathrm{s}\hspace{-0.8ex}.795,\ \delta\ =\ -37^\circ54\mathrm{'}35\mathrm{''}\hspace{-0.8ex}.54,\\
	\pi &=& \unit[8{.}16\pm0{.}80]{mas},\\
	\mu_{\alpha}^* &=& \unit[325{.}9\pm2{.}3]{mas/yr},\\
	\mu_{\delta} &=& \unit[-59{.}2\pm2{.}1]{mas/yr},
	\end{array}\label{eq:1856input}
\end{equation}

\noindent where $\mu_{\alpha}^*$ is the proper motion in right ascension corrected for declination.\\
Given its proper motion and parallax, the transverse velocity of \rxja{} is $v_t=\unit[192^{+17}_{-21}]{km/s}$. \citet{2001A&A...380..221V} tried to measure the inclination $i$ of the bow shock which \rxja{} creates in the interstellar medium (ISM). Unfortunately, they could not obtain a precise result but estimated $i=\unit[60\pm15]{deg}$ or $i$ even closer to $\unit[90]{deg}$, depending on the model. A lower limit on $i$ of $\unit[45]{deg}$ implies a maximum radial velocity modulus of $\approx\unit[250]{km/s}$ ($3\sigma$\,$v_t$). We will address this issue later.\\

\begin{figure}
\centering
\includegraphics[width=0.25\textwidth, viewport= 40 205 540 600]{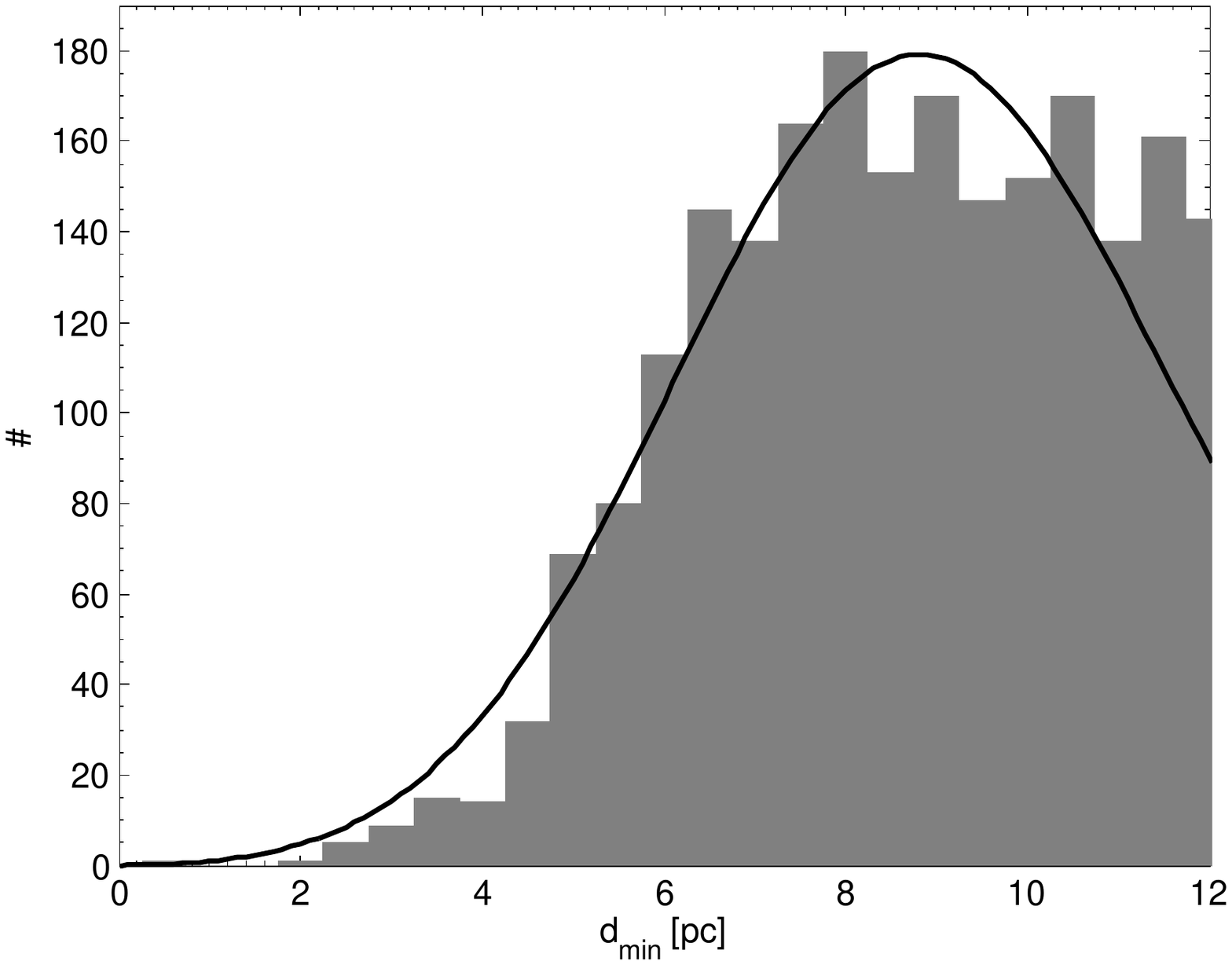}\includegraphics[width=0.25\textwidth, viewport= 40 205 540 600]{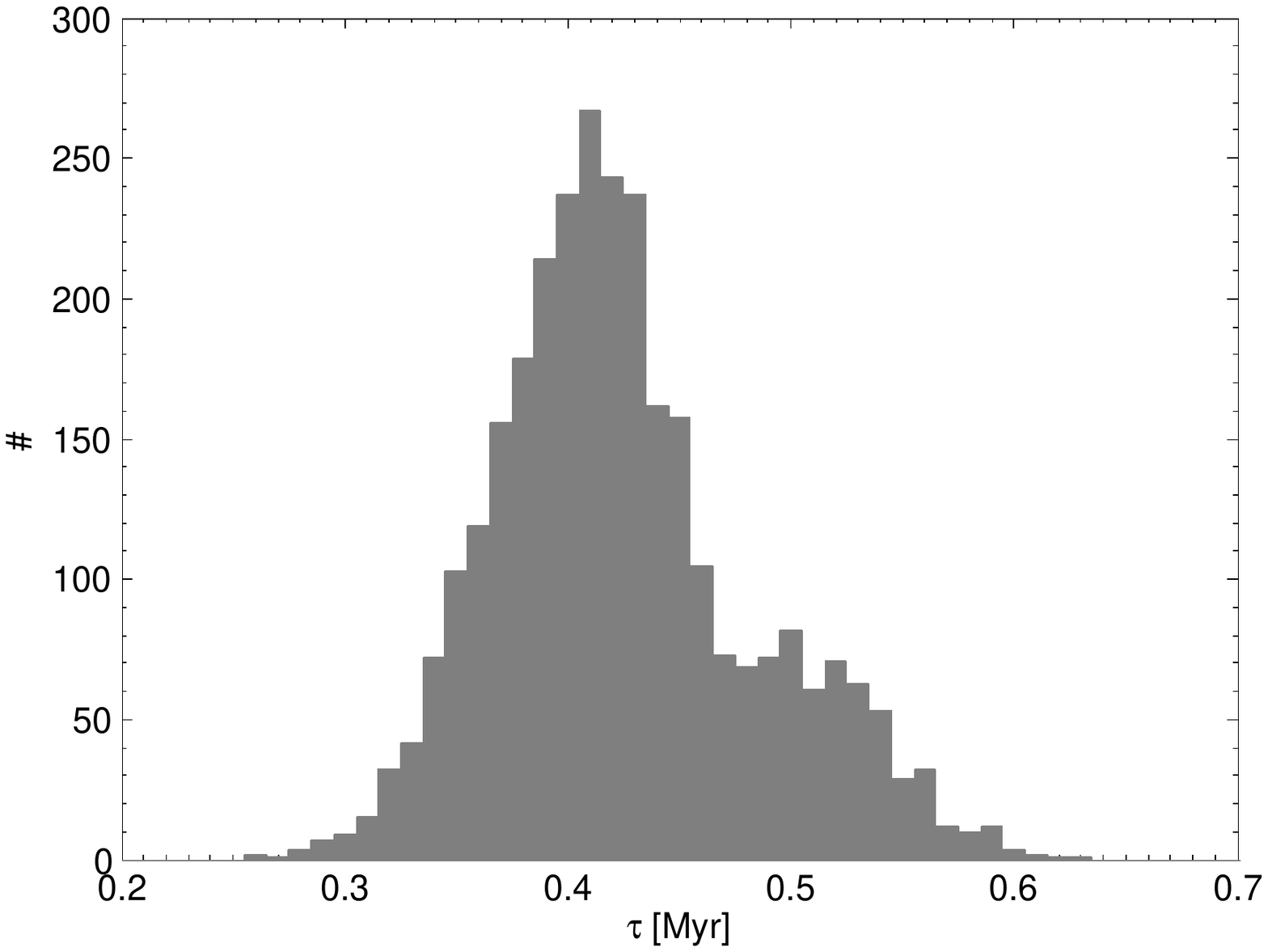}\linebreak
\includegraphics[width=0.25\textwidth, viewport= 40 205 540 600]{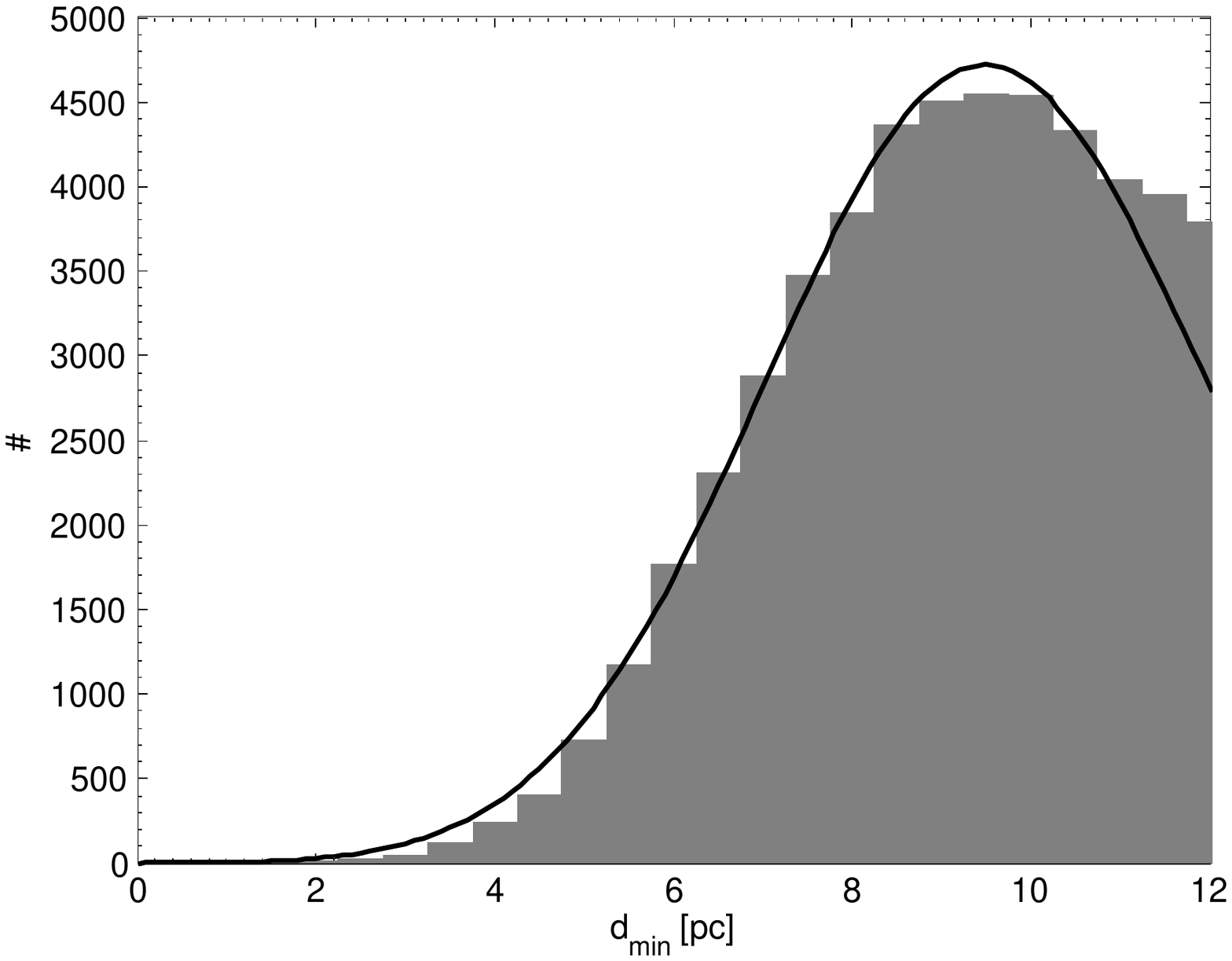}\includegraphics[width=0.25\textwidth, viewport= 40 205 540 600]{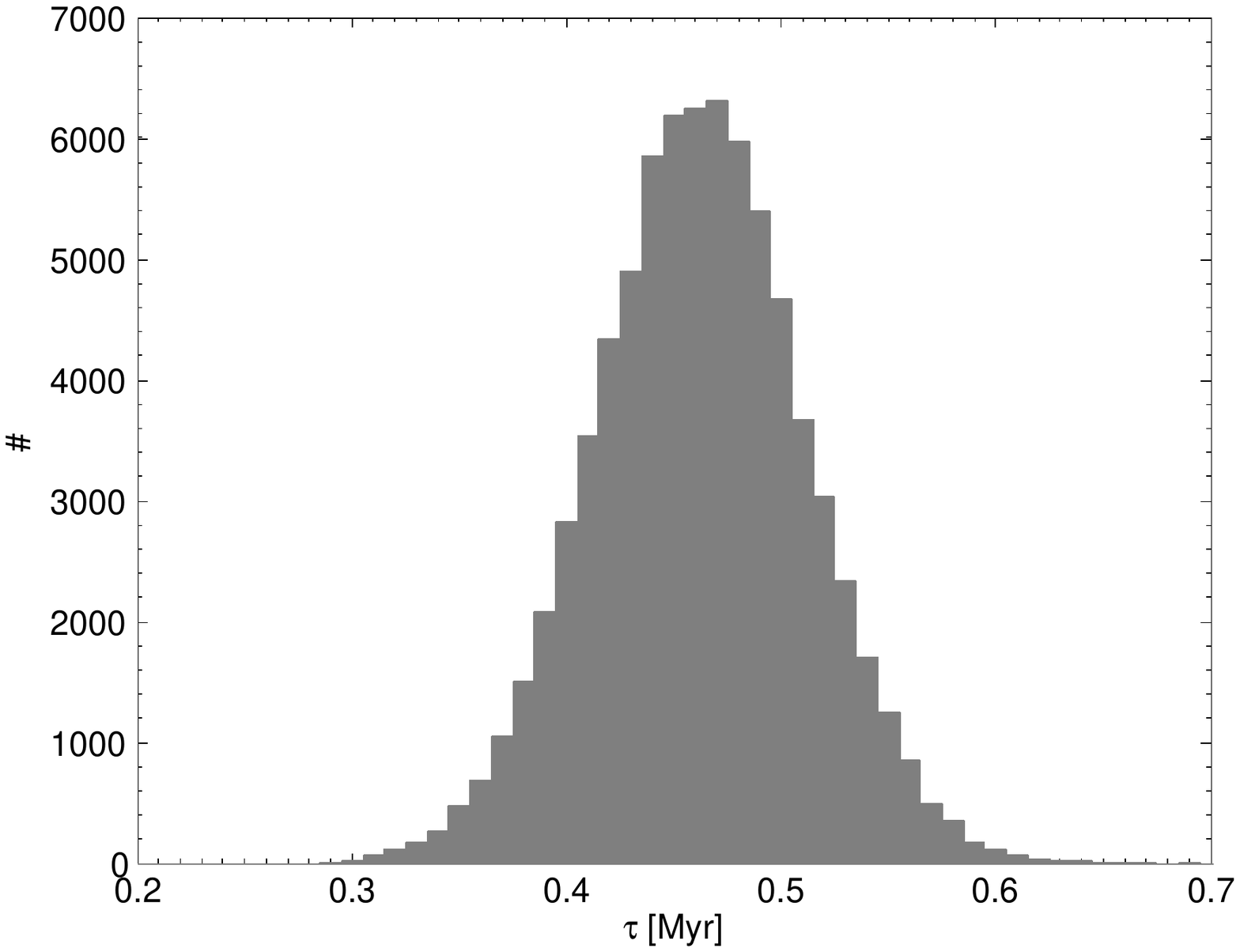}
\caption{Left: Distribution of minimum separations $d_{min}$ between \rxja{} and the US centre. Drawn as well is the theoretically expected curve. Right: Distribution of corresponding flight times $\tau$ in the past since the SN.\newline
Top panel: Distributions for $v_r$ distribution derived from the \citet{2005MNRAS.360..974H} distribution of pulsar space velocities. Solid curve in the left panel: $\mu=7{.}9$, $\sigma=2{.}0$. Bottom panel: Distributions for $v_r=-250\ldots+250\,\mathrm{km/s}$. Solid curve in the left panel: $\mu=8{.}8$, $\sigma=1{.}8$.\newline
This indicates that the closest approach between \rxja{} and the US centre was $\unit[8-9]{pc}$ almost half a million years ago.}\label{fig:hists1856US}
\end{figure}

First, we perform $10^4$ Monte Carlo runs to find close encounters between \rxja{} and any association/cluster in the past five million years\footnote{Comparing their effective temperatures with cooling curves, \rxja{} and \rxjn{} cannot be older \citep[see e.g.][]{2004ApJS..155..623P,2004Sci...304..536L}.}. We select those associations/clusters for which the smallest separation $d_{min}$ found was less than three times the association/cluster radius, 18 in total. For those 18 associations/clusters, we carry out another one million Monte Carlo runs. For 13 of them we find close encounters consistent with the association/cluster boundaries; however, most of them [Tucana/Horologium (Tuc-Hor), the $\beta$ Pictoris group ($\beta$ Pic-Cap), AB Doradus (AB Dor), Hercules-Lyrae, Sagittarius OB5, Scorpius OB4 (Sco OB4), Pismis 24, Trumpler 27, NGC 6383, van den Bergh-Hagen 217, NGC 6396] can be excluded because in these cases the radial velocity would need to be $\left|v_r\right|\gtrsim\unit[350]{km/s}$ (larger than maximum $v_r$ inferred from bow shock, cf. above discussion on the bow shock). \\
We adapt the theoretically expected distribution to the first part of the $d_{min}$ distribution such that the slope and peak can be adjusted (see also \autoref{fig:hists1856US} for an example). Also, in all but one (Sco OB4) of these cases, adapting the theoretical curve to the $d_{min}$ distribution suggests that the closest approach of \rxja{} to the association/cluster was outside the association/cluster boundaries. For the extended Corona-Australis association (Ext. R CrA), small separations between \rxja{} and the association centre are found for $\unit[0{.}04\pm0{.}02]{Myr}$ in the past. This is not surprising because the present position of \rxja{} lies within the association and it would need up to $\approx\unit[0{.}2]{Myr}$ to cross the association (assuming a maximum space velocity of $\unit[350]{km/s}$, i.e. $3\sigma\,v_t$ and $v_{r,max}=\unit[250]{km/s}$, and an extension of Ext. R CrA of $\unit[62]{pc}$). \\
Hence, Ext. R CrA is not the parent association of \rxja{} unless the NS is only a few $\unit[10000]{yr}$ old, however then we would probably still see the SN remnant (SNR) (there is no SNR known in this area, \citealt{2009BASI...37...45G}; A. Poghosyan, priv. comm.). The only association that fits perfectly to being the parent association of \rxja{} is Upper Scorpius (US) (as suggested before by \citealt{2002ApJ...576L.145W} using a much simpler calculation and T10 using a larger distance). \autoref{tab:US_prop} (a) summarises the derived NS parameters and the position and time of the SN (for the deduction of the properties see T10 or additional online supporting information). The displacement between \rxja{} and the US centre would be $\unit[7{.}9]{pc}$ as inferred from the theoretically expected curve. The $d_{min}$ distribution and the distribution of the corresponding flight times $\tau$ are shown in \autoref{fig:hists1856US}, top panel. The SN would have occurred well within US that has a radius of $\approx\unit[15]{pc}$. We find that the radial velocity of \rxja{} is probably very small (now indicated by both the parent association and the bow shock) although utilising the velocity distribution for pulsar velocities gives priority to larger values. \\
Since both the radial velocity and the parallax are strongly correlated as a larger distance can be compensated for by a larger radial velocity, parallaxes may be slightly biased towards smaller values (since larger radial velocities are preferred in the calculations). Therefore, we repeat the simulation now assuming a uniform distribution of the radial velocity of \rxja{} in the range of $v_r=\unit[-250\ldots+250]{km/s}$ (see above discussion on the bow shock; note that we do not give priority to $v_r=\unit[0]{km/s}$). The results are summarised in \autoref{tab:US_prop} (b). Indeed, the parallax turns out to be larger (distance smaller) and the radial velocity tends to be close to zero. The derived values [\autoref{tab:US_prop} (b)] for the proper motion and radial velocity imply an inclination of $i=\unit[88\pm6]{deg}$, consistent with the observation of the bow shock \citep{2001A&A...380..221V}. The displacement between \rxja{} and the US centre would be $\unit[8{.}8]{pc}$ as inferred from the theoretically expected curve. The $d_{min}$ distribution and the distribution of the corresponding flight times $\tau$ are shown in \autoref{fig:hists1856US}, bottom panel.\\
Assuming contemporary star formation, the mass of the progenitor star of \rxja{} is given as the age of US ($\unit[5\pm2]{Myr}$, e.g. \citealt{2010A&A...522A..51D,2002AJ....124..404P,2001AJ....121.1040P,1999AJ....117.2381P,1978ppeu.book..101B,1989A&A...216...44D}) minus the time since the potential SN ($\approx\unit[0{.}5]{Myr}$). Using evolutionary models from \citet{1980FCPh....5..287T}, \citet{1989A&A...210..155M} and \citet{1997PhDT........31K} we estimate it to be $\unit[45\pm3]{M_\odot}$, $\unit[43^{+21}_{-9}]{M_\odot}$ and $\unit[37^{+27}_{-9}]{M_\odot}$, respectively, corresponding to a main-sequence spectral type of O6, O5 to O7 and O5 to O7 for the respective model \citep{Schmidt-Kaler1982}. The progenitor star of \rxja{} should have an earlier spectral type than the earliest present member which has spectral type B0 (HIP 81266, \citealt{2002A&A...381..446M}). Although NSs are believed to form from progenitors with masses of $\lesssim\unit[20-25]{M_\odot}$ \citep[e.g.][]{2003ApJ...591..288H} it is known that in binary systems also more massive stars, $\gtrsim\unit[40]{M_\odot}$, may produce NSs \citep[e.g.][]{2008ApJ...685..400B,2002ApJ...578..335F,1984Natur.309..598V}. Then, it should have ejected a runaway star, which may be known or unknown, yet.
\begin{table}
\centering
\caption{Predicted current parameters of \rxja{} if it was born in US using as input for the radial velocity (a) a probability distribution according to \citet{2005MNRAS.360..974H} and (b) a uniform distribution in the range of $v_r=\unit[-250\ldots+250]{km/s}$; $v_{sp}$ denotes the space velocity. The time and position of the SN is given as well. For the deduction of the radial velocity $v_r$, parallax $\pi$, proper motion $\left(\mu_\alpha^*,\mu_\delta\right)$, the distance of the SN from the Sun $d_\odot$ (at the time of the SN), the SN equatorial coordinates ($\alpha$, $\delta$, J2000.0, as seen from the Earth at present) and the time in the past at which the SN occurred ($\tau$, the kinematic age of the NS), see additional online supporting information. The distance of the SN to the US centre was obtained by adapting the theoretically expected curve (equation 1 in T10) to the slope of the $d_{min}$ distribution.}\label{tab:US_prop}
\begin{tabular}{l c c}
\toprule
\multicolumn{3}{c}{Predicted present day parameters of \rxja{}}\\\midrule
												 & (a)								& (b)\\
$v_r$ [km/s]             & $29^{+28}_{-20}$		&	$6^{+19}_{-20}$\\
$\pi$ [mas]              & $7{.}0^{+0{.}6}_{-0{.}2}$	& $8{.}1^{+0{.}4}_{-0{.}6}$\\
$\mu_{\alpha}^*$ [mas yr$^{-1}$] & $325{.}9\pm2{.}3$	& $325{.}9\pm2{.}3$\\
$\mu_\delta$ [mas yr$^{-1}$]     & $-59{.}4\pm2{.}1$	& $-59{.}3\pm2{.}1$\\
$v_{sp}$ [km/s]					 & $219^{+16}_{-12}$					&	$195^{+10}_{-14}$\\\midrule
\multicolumn{3}{c}{Predicted SN position}\\\midrule
distance to the Sun [pc] & $151^{+5}_{-5}$	& $156^{+3}_{-8}$\\
right ascension [deg]    & $241{.}8^{+0{.}7}_{-1{.}3}$ 	& $241{.}4^{+1{.}8}_{-1{.}1}$\\
declination [deg]        & $-25{.}5^{+0{.}7}_{-0{.}4}$	& $-25{.}7^{+0{.}7}_{-0{.}9}$\\
time in the past [Myr]   & $0{.}41^{+0{.}05}_{-0{.}06}$	& $0{.}46^{+0{.}05}_{-0{.}05}$\\
distance from US centre [pc] & $7{.}9\pm2{.}8$	& $8{.}8\pm2{.}5$\\\bottomrule
\end{tabular}
\end{table}

\subsection{Searching for a former companion of \rxja{}}\label{subsec:1856comp}

From \autoref{subsec:1856parentassoc}, US should be the parent association of \rxja{}. Now, our attempt is to find a runaway star that might have been the former companion of the NS's progenitor.\\
With US hosting the birth place of \rxja{}, we know that the radial velocity of the NS is very small ($v_r=\unit[6^{+19}_{-20}]{km/s}$). For the following analysis we adopt $v_r=\unit[0\pm50]{km/s}$.\\
Similar to \autoref{subsec:1856parentassoc}, we calculate the trajectories of \rxja{} and any young runaway star with full 3D kinematics from T11, 1705 stars in total. \\
First, we perform $10^3$ runs to find runaway stars which might have crossed the path of \rxja{} in the past. We find 16 stars for which the smallest separation to the NS found was less than $\unit[10]{pc}$ (see additional online supporting information). Given their ages and kinematics, only three of them, HIP 74717, HIP 76028 and HIP 78681, could have originated from the US association ($\unit[0{.}5^{+0{.}1}_{-0{.}1}]{Myr}$, $\unit[1{.}8^{+0{.}6}_{-0{.}2}]{Myr}$ and $\unit[0{.}1^{+0{.}1}_{-0{.}1}]{Myr}$ ago, respectively) and only for them close encounters with \rxja{} are found to be situated inside US. After further three million runs for these three stars, we exclude HIP 76028 since the smallest separation found was larger than one parsec, see additional online supporting information for details on the selection criteria; not surprisingly since we find that the star probably originated from or at least passed US $\approx\unit[1]{Myr}$ before \rxja{} was born; if it is a BSS runaway star it may have formed in another SN in the Scorpius-Centaurus associations (14 to 20 SNe already exploded, \citealt{2009SSRv..143..437F}).\\ 

For HIP 74717 the smallest $d_{min}$ found was $\unit[0{.}06]{pc}$. Adapting the theoretical distribution for 3D separations to the slope of the $d_{min}$ distribution (restricting to 7558 runs where both, the runaway star and \rxja{} were within the US boundaries, $R_{US}=\unit[15]{pc}$) implies a fly-by of the runaway and \rxja{} with a displacement to each other of $\unit[7{.}3\pm2{.}8]{pc}$ $\unit[0{.}45^{+0{.}03}_{-0{.}02}]{Myr}$ in the past. Moreover, HIP 74717 is a single-lined spectroscopic binary with a rotational velocity of $\unit[38]{km/s}$ \citep{2002A&A...384..491C}, hence probably not a BSS runaway star. \\
In the case of HIP 78681, the smallest $d_{min}$ value found was $\unit[0{.}04]{pc}$. In only 329 Monte Carlo runs the distance of both, the runaway star and \rxja{}, to the US centre was less than $\unit[15]{pc}$, the nominal radius of the association. These runs yield $d_{min}$ values (i.e. separations between the NS and the runaway star) from $\approx0$ to $\approx\unit[25]{pc}$. To improve statistics to be able to compare the $d_{min}$ histogram with the theoretically expected curve, we include runs for which both stars were within $\unit[20]{pc}$ from the US centre, i.e. in a region at the US border. The number of runs rises to 49277. Then, the number of runs yielding large separations $d_{min}$ is large; however, the peak at very small $d_{min}$ values becomes well visible. For the first bins of this histogram, the theoretical curve suggests that both objects could have been very close to each other (positional difference $\unit[0\pm1{.}1]{pc}$) $\unit[0{.}53^{+0{.}04}_{-0{.}05}]{Myr}$ in the past. In the Simbad\footnote{http://simbad.u-strasbg.fr/simbad/} database, HIP 78681 is listed with spectral type G7II whereas \citet{2004AA...424..727P} list it as barium star (G8IIIBa1). Moreover, HIP 78681 is a single-lined spectroscopic binary. Hence, the system is probably old with the companion of HIP 78681 being a white dwarf as widely excepted for SB1 barium stars \citep{1984ApJ...278..726B}. However, if HIP 78681 is a young BSS runaway star, it might be the former companion to \rxja{} (although it is difficult to judge since relaxation of the US boundary was necessary).\\
Thus, we do not find a convincing runaway star in the sample of runaway stars with full kinematics from T11 to be a suitable former companion candidate for \rxja{}.\\
Note, that the classical runaway star HIP 81377 (=$\zeta$ Ophiuchi) is again not found to be the former companion to \rxja{} as it was suggested by \citet{2000IAUS..195..437W} but excluded by \citet{2001A&A...365...49H} and T10 (we found a smallest separation of $\unit[18{.}8]{pc}$ to the US centre after three million Monte Carlo runs). \\

Since there are also 842 runaways in the runaway star catalogue (T11) without radial velocities, we examined whether one of those could have been close to \rxja{} in the past. We varied the radial velocity for those stars randomly within $\pm\unit[500]{km/s}$ (the largest radial velocity values among all catalogue stars are $\approx\unit[\pm400]{km/s}$). After $10^3$ Monte Carlo runs 24 stars were found for which close encounters ($\leq\unit[10]{pc}$, see additional online supporting information) might have been possible in the past five million years. Nine of them showed $d_{min}$ values smaller than $\unit[1]{pc}$ after $10^5$ runs, hence were chosen for a more detailed investigation with $10^6$ runs (for justification of the chosen limits see additional online supporting information).\\
The distribution of the peculiar spatial velocities $v_{sp,run}$ of the population of young runaway stars is well represented by a Maxwellian distribution with a velocity dispersion of $\sigma=\unit[24{.}4]{km/s}$ (T11) and a maximum of $\unit[34{.}5]{km/s}$. Extraordinary high velocities are unlikely although possible for individual cases. Only three stars -- HIP 63803, HIP 70438 and HIP 74219 -- would need plausible spatial velocities for close encounters with \rxja{}; for the other stars, the necessary spatial velocity deviates from the distribution maximum by more than $6\sigma$. In \autoref{tab:4cand} the properties of the potential close encounter between \rxja{} and each of the three runaway stars are given. \\
Considering the large encounter separation in the case of HIP 70438, it is unlikely that HIP 70438 could have been at the same place as \rxja{} in the past. \\
In the case of HIP 63803 the position of the close encounter would have been far outside US. The spectral type of HIP 63803 is K2 (Simbad) and its position in the Hertzsprung-Russel diagram (HRD) suggests it to be a giant with an age of $\approx\unit[50-90]{Myr}$ (T11). Provided that a potential companion to that star formed at the same time and comparing to the proposed encounter time of $\approx\unit[1{.}5]{Myr}$ in the past, the progenitor of \rxja{} should have been at least as old as $\unit[48]{Myr}$ when it exploded in a SN. Stars that end their lives in SNe do not live that long. Hence, HIP 63803 cannot be the former companion to \rxja{} and might not be a BSS runaway star.\\
In T11 the F5V star HIP 74219 is listed with a mass of $\unit[1{.}2\pm0{.}1]{M_\odot}$ and an age of $\unit[39\pm19]{Myr}$ that were determined from the median value of different evolutionary models not taking into account the error on the luminosity (mainly due to the parallax uncertainty) of the star. In the HRD the star lies just below the ZAMS of the evolutionary models. For that reason, it was treated as ZAMS star in T11. For higher luminosities according to its luminosity error the models predict ages up to a few Gyr. Hence, it seems unlikely that HIP 74219 was ejected from the young US association ($\approx\unit[5]{Myr}$), but we cannot fully exclude it as the former companion of \rxja{}.\\
So, we also do not find a convincing runaway star in the sample of runaway stars without radial velocity measurements from T11 to be a suitable former companion candidate for \rxja{}.
\begin{table}
\caption{Predicted properties of potential close encounters between \rxja{} and the three runaway stars without known radial velocity measurements. The first column gives the Hipparcos identifier. Columns 2 to 5 give the distance of the encounter to the centre of US ($d_{\mathrm{US}}$), the separation $d$ inferred from the theoretical curve ($\mu\pm\Delta$), the time $\tau$ of the encounter in the past as well as the peculiar spatial velocity $v_{sp,run}$ of the runaway star needed.}\label{tab:4cand}
\begin{tabular}{c c c c c}
\toprule
HIP			&  $d_{\mathrm{US}}$& $d$						& $\tau$												& $v_{sp,run}$\\
				&	 [pc]						& [pc]						& [Myr]													& [km/s]\\\midrule
63803		&  $240\ldots270$	& $0{.}0\pm12{.}0$	& $1{.}42^{+0{.}15}_{-0{.}10}$	& $173^{+95}_{-52}$\\
70438		&	 $50\ldots70$		&	$6{.}7\pm3{.}4$	&	$0{.}67^{+0{.}10}_{-0{.}09}$	& $10^{+158}_{-2}$\\
74219		&	 $0\ldots35$		&	$0{.}0\pm4{.}0$ &	$0{.}55^{+0{.}08}_{-0{.}04}$	& $66^{+22}_{-18}$\\
\bottomrule
\end{tabular}
\end{table}

\subsection{Identifying the parent association of \rxjn{}}\label{subsec:0720parentassoc}

Potential parent associations of \rxjn{} are investigated in the same way as for \rxja{} (\autoref{subsec:1856parentassoc}). We repeat the investigations already carried out in T10 here again as a new parallax measurement was done by \citet{ThomasPhD}, yielding a distance of $\unit[280^{+210}_{-85}]{pc}$. Compared to the old value of $\unit[360^{+170}_{-90}]{pc}$ \citep{2007ApJ...660.1428K}, this new distance is in much better agreement with estimations derived from the spectrum and hydrogen column density $n_H$ \citep{2007Ap&SS.308..171P} giving $\unit[250\pm25]{pc}$ although it is consistent within the error bars with the older value.\\
In contrast to \rxja{}, the radial velocity of \rxjn{} is unconstrained (no bow shock detected) and is therefore derived from the probability distribution of pulsar space velocities by \citet{2005MNRAS.360..974H}. We adopt the following parameters for \rxjn{}:

\begin{equation}
	\begin{array}{l c l}
	\alpha &=& 07^\mathrm{h}20^\mathrm{m}24^\mathrm{s}\hspace{-0.8ex}.961,\ \delta\ =\ -31^\circ25\mathrm{'}50\mathrm{''}\hspace{-0.8ex}.21\\
				 & & \mbox{(\citealt{2003ApJ...590.1008K})},\\
	\pi &=& \unit[3{.}6\pm1{.}6]{mas}\ \mbox{(\citealt{ThomasPhD})},\\
	\mu_{\alpha}^* &=& \unit[-92{.}8\pm1{.}4]{mas/yr}\ \mbox{(\citealt{ThomasPhD})},\\
	\mu_{\delta} &=& \unit[55{.}3\pm1{.}7]{mas/yr}\ \mbox{(\citealt{ThomasPhD})}.
	\end{array}\label{eq:0720input}
\end{equation}
\\

For \rxjn{}, 18 associations/clusters are found for which separations $d_{min}$ are consistent with the associations/cluster boundaries after $10^6$ Monte Carlo runs. We then adapt equation 1 in T10 to the first bins of each $d_{min}$ distribution to obtain the distance $d$ of the SN to the association centre. Comparing the radii of each association with this putative separation $d$ of the SN from the association centre, nine associations/clusters are found to be potential birth places of \rxjn{}, i.e. $d$ is consistent with the association boundaries within its standard deviation: TWA, Tuc-Hor, $\beta$ Pic-Cap, the HD 141569 group, AB Dor, Collinder 140 (Col 140), Tr 10 and the Carina (CarA) and Argus Associations. In \autoref{tab:assoc0720_2} we give the position of the SN and the properties \rxjn{} would currently have if it was born in the respective association. \\
Note that for the HD 141569 group, AB Dor and Tr 10 the proposed SN position is near the edge of the respective association. The HD 141569 group may be excluded from the list of potential parent associations since it contains only three stellar systems (five stars in total, HD 141569 is a triple system itself, \citealt{2001ASPC..244...75W}). Here, we retain it in the list of potential birth places (\autoref{tab:assoc0720_2}) for completeness.\\
AB Dor currently contains young dwarfs with spectral types ranging from mid-F to early M (earliest members: HIP 18859 and HIP 19183 with F5, \citealt{2004ApJ...613L..65Z,2006ApJ...643.1160L}) and masses from $0{.}7$ to $\unit[1{.}3]{M_\odot}$ [derived from their positions in the HRD using evolutionary tracks (see T11 for references)]. Hence, it seems unlikely that a SN occurred in ABDor although a $\unit[10]{M_\odot}$ star might be expected to have formed in AB Dor from the comparison of its mass function with the initial mass function \citep{2005ASSL..327..175K}.\\
Tr 10 is listed as sparse open cluster with a diameter of $\approx\unit[4]{pc}$ (29' at $\unit[424]{pc}$) in the Open Cluster Catalogue \citep{2002A&A...389..871D}. However, \citet{1999AJ....117..354D} and \citet{1999osps.conf..411B} found it to be actually an association with a diameter of $\approx\unit[50]{pc}$ ($8^\circ$ at $\unit[366]{pc}$) rather than a cluster. The 23 members found by \citet{1999AJ....117..354D} are all B3 to A0 type stars. Since early B type stars are present in Tr 10, it is plausible that the association already experienced a SN.\\
Close encounters between \rxjn{} and Col 140 occur at $\unit[0{.}04\pm0{.}01]{Myr}$ in the past. In this case, the SNR should still be visible; however there is no known SNR in this area (\citealt{2009BASI...37...45G}; A. Poghosyan, priv. comm.).\\
\begin{table*}
\centering
\caption{Potential parent associations of \rxjn{}.\newline
In the second and third Columns we give the predicted encounter separation $d$ (inferred from the theoretical curve) and encounter time $\tau$ (determined from those runs which yield separations consistent with $d$). Columns 4 to 8 give the predicted present NS parameters (radial velocity $v_r$, proper motion $\mu_\alpha^*$ and $\mu_\delta$, parallax $\pi$) and space velocity $v_{sp}$ for each case and Columns 9 to 11 indicate the distance to the Sun $d_\odot$ (at the time of the SN) and equatorial coordinates (J2000.0, as seen from Earth at present) of the potential SN. Error bars denote 68 per cent confidence (cf. Appendix B, T10).}
\label{tab:assoc0720_2}
\setlength\extrarowheight{3pt}
\small
\begin{tabular}{c o{4.2} >{$}r<{$} >{$}r<{$} o{4.2} o{4.2} >{$}r<{$} >{$}r<{$} >{$}c<{$} >{$}r<{$} >{$}r<{$}}
\toprule
Assoc.	&	\multicolumn{1}{c}{$d$} & \multicolumn{1}{c}{$\tau$} &	\multicolumn{5}{c}{Predicted current NS parameters}	&	d_{\odot} 	& \multicolumn{1}{c}{$\alpha$}	& \multicolumn{1}{c}{$\delta$}\\ 
	&	&  &	\multicolumn{1}{c}{$v_r$}			& \multicolumn{1}{c}{$\mu_{\alpha}^*$} & \multicolumn{1}{c}{$\mu_{\delta}$} & \multicolumn{1}{c}{$v_{sp}$} &	\multicolumn{1}{c}{$\pi$}						&		& & \\ 
		& \multicolumn{1}{c}{[pc]} & \multicolumn{1}{c}{[Myr]}	& \multicolumn{1}{c}{[km/s]} & \multicolumn{1}{c}{[mas/yr]} & \multicolumn{1}{c}{[mas/yr]} & \multicolumn{1}{c}{[km/s]} & \multicolumn{1}{c}{[mas]} & \multicolumn{1}{c}{[pc]} & \multicolumn{1}{c}{[$^\circ$]} & \multicolumn{1}{c}{[$^\circ$]}\\\midrule
TWA					& 0{.}0+2{.}4	& 0{.}41^{+0{.}09}_{-0{.}06}	& 376^{+156}_{-28} & -92{.}8+1{.}4 & 55{.}3+1{.}7 & 416^{+110}_{-74} & 4{.}4^{+0{.}5}_{-0{.}5}	& 58^{+2}_{-5}			& 178{.}1^{+3{.}4}_{-2{.}5}	& -41{.}2^{+0{.}8}_{-1{.}0}\\
Tuc-Hor			& 45{.}6+2{.}7& 0{.}28^{+0{.}04}_{-0{.}04}	& 529^{+91}_{-59}& -92{.}8+1{.}4	& 55{.}3+1{.}7	& 540^{+70}_{-85} & 5{.}4^{+0{.}6}_{-0{.}4}& 33^{+3}_{-4}			& 167{.}8^{+6{.}2}_{-6{.}8}	& -42{.}8^{+0{.}8}_{-0{.}8}\\
$\beta$ Pic-Cap	& \multicolumn{1}{c}{$34^{+8}_{-8}$}	& 0{.}44^{+0{.}01}_{-0{.}11}& 491^{+119}_{-71}& -92{.}8+1{.}4	& 55{.}3+1{.}7	& 501^{+89}_{-71} & 5{.}1^{+0{.}7}_{-0{.}5}& 44^{+8}_{-8}			& 205{.}2^{+4{.}4}_{-3{.}8}	& -32{.}8^{+1{.}8}_{-2{.}4}\\
HD 141569		& 16{.}5+1{.}7	& 0{.}61^{+0{.}19}_{-0{.}07}	& 396^{+107}_{-41}	& -92{.}9+1{.}4	& 55{.}1+1{.}7	& 424^{+81}_{-65} & 4{.}1^{+0{.}5}_{-0{.}5}& 102^{+8}_{-3}	& 245{.}2^{+2{.}2}_{-3{.}4}	& -6{.}4^{+3{.}4}_{-1{.}6}\\
AB Dor			& \multicolumn{1}{c}{$55^{+13}_{-16}$}	& 0{.}37^{+0{.}07}_{-0{.}06}		& 478^{+110}_{-60}	& -92{.}8+1{.}4	& 55{.}3+1{.}7	& 491^{+87}_{-81} & 4{.}8^{+0{.}8}_{-0{.}3}& 36^{+10}_{-9}	& 207{.}8^{+4{.}4}_{-4{.}3}	& -32{.}3^{+2{.}6}_{-1{.}9}\\
Col 140			& 1{.}8+0{.}8	& 0{.}05^{+0{.}05}_{-0{.}05}		& -670\ldots+590	& -92{.}8+1{.}4	& 55{.}3+1{.}7	& 463^{+119}_{-192} & 2{.}8^{+0{.}1}_{-0{.}2}& 375^{+2}_{-2}	& 111{.}2^{+0{.}2}_{-0{.}2}	& -32{.}0^{+0{.}2}_{-0{.}1}\\
Tr 10				& 22{.}9+2{.}7	& 0{.}50^{+0{.}05}_{-0{.}05}		& 274^{+151}_{-43}	& -92{.}7+1{.}4	& 55{.}6+1{.}6	& 390^{+112}_{-60} & 1{.}9^{+0{.}2}_{-0{.}1}& 373^{+10}_{-9}	& 133{.}5^{+0{.}9}_{-1{.}1}	& -39{.}6^{+0{.}4}_{-0{.}4}\\
CarA				& 33{.}9+1{.}1	& 0{.}34^{+0{.}06}_{-0{.}06}		& 404^{+146}_{-74}	& -92{.}7+1{.}4	& 55{.}5+1{.}6	& 427^{+118}_{-104} & 4{.}1^{+0{.}9}_{-0{.}4}& 79^{+7}_{-3}	& 139{.}6^{+4{.}4}_{-4{.}2}	& -41{.}5^{+1{.}0}_{-0{.}6}\\
Argus 			& 35{.}1+1{.}1	& 0{.}35^{+0{.}08}_{-0{.}08}			& 388^{+158}_{-65}	& -92{.}7+1{.}4	& 55{.}6+1{.}6	& 390^{+135}_{-90} & 4{.}0^{+0{.}5}_{-0{.}6}& 103^{+6}_{-3}	& 134{.}9^{+4{.}8}_{-2{.}4}	& -40{.}4^{+0{.}7}_{-0{.}9}\\
\bottomrule
\end{tabular}
\end{table*}
Hence, except the HD 141569 group and the open cluster Col 140, the associations listed in \autoref{tab:assoc0720_2} are still probable birth associations of \rxjn{} although AB Dor seems unlikely.

\subsection{Searching for a former companion of \rxjn{}}\label{subsec:0720comp}

\begin{figure}
\centering
\includegraphics[width=0.4\textwidth, viewport= 40 205 540 600]{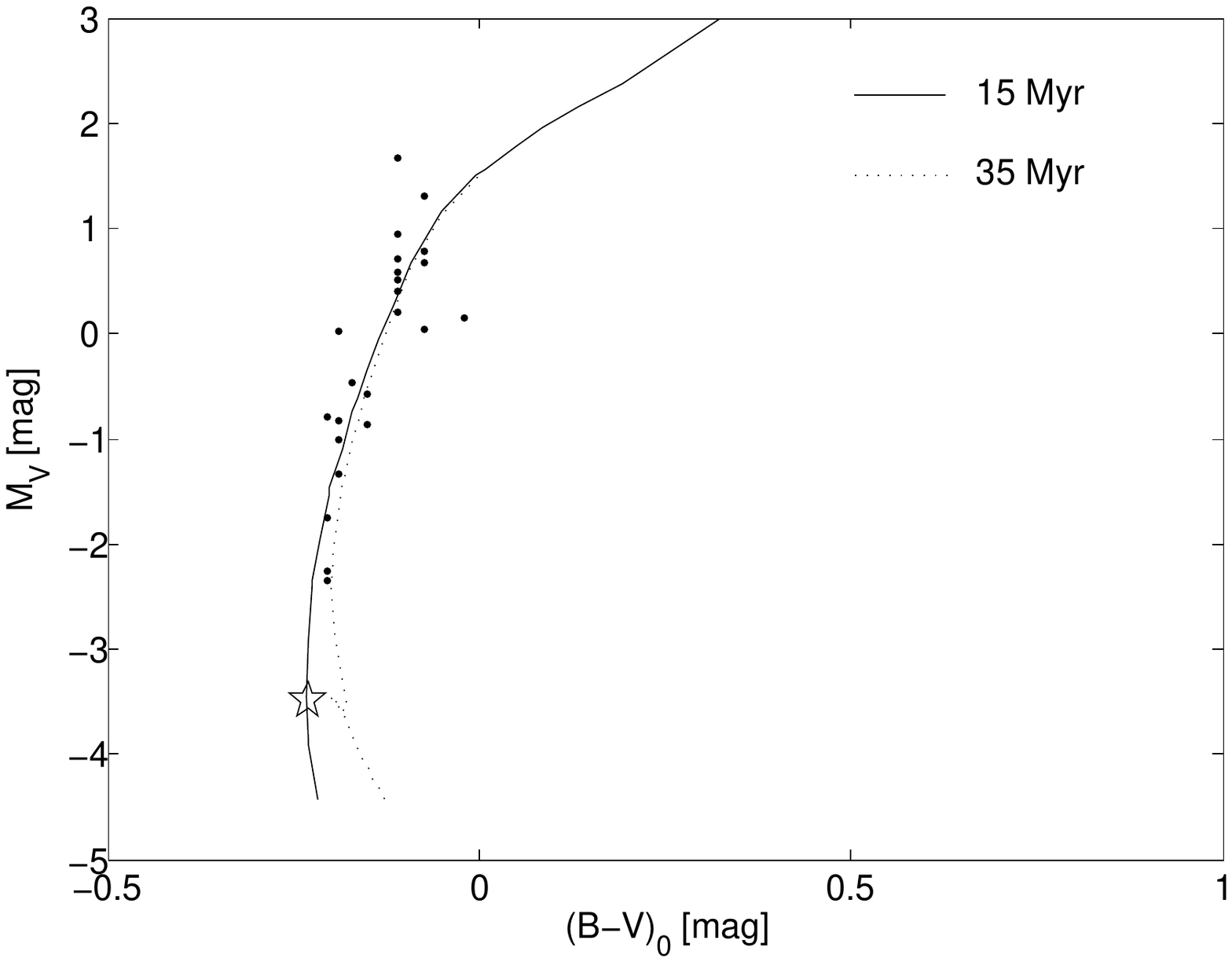}
\caption{$\left(B-V\right)_0$ versus $M_V$ diagram of Tr 10 (dots represent member stars according
to \citealt{1999AJ....117..354D}, the star marks HIP 43158). The solid and dashed lines represent the $\unit[15]{Myr}$ and $\unit[35]{Myr}$ isochrones from \citet{2008A&A...482..883M} (for solar metallicity; http://stev.oapd.inaf.it/cgi-bin/cmd). $\left(B-V\right)_0$ and $M_V$
of the stars are derived from their spectral types according to Schmidt-Kaler (1982).}
\label{fig:Tr10HRD}
\end{figure}

After $10^3$ Monte Carlo runs 53 runaway stars with full kinematics show a smallest separation $d_{min}$ to \rxjn{} of less than $\unit[10]{pc}$. For them, another three million runs are performed (see additional online supporting information). We select those stars for which separations smaller than one parsec are found after three million runs (see additional supporting information), 29 in total.\\
To further reduce the number of potential former companion candidates, we investigate whether the potential close encounters occurred inside one of the possible birth associations of \rxjn{} listed in \autoref{tab:assoc0720_2}. Possible close encounters between \rxjn{} and 16 stars are found to be consistent with the radius of one or more of the nine associations/clusters.\\
Of them, we can exclude HIP 36160 (spectral type G5V) whose chromospheric age is $\unit[3{.}27]{Gyr}$ \citep{2004AA...423..517R}, the cataclysmic variable HIP 40430 (spectral type O9nne) and the barium star HIP 47267 (\citealt{1997AA...321L...9B}; spectral type G8II). The latter has been included in the sample of T11 owing to its luminosity class. We also exclude HIP 63972 (spectral type K0II-III) that has been classified as old disk star \citep{1993AJ....106...80E} and entered the sample of T11 due to its luminosity class. The ages for HIP 67655 (spectral type G5V), HIP 68783 (G5), HIP 74689 (spectral type A4V) and HIP 79377 (spectral type G1/G2V) are very uncertain since these stars currently lie on or close to the main-sequence. From its position in the HRD we estimate the age of HIP 74689 to be roughly $50$ to $\unit[300]{Myr}$, too old for an origin in $\beta$ Pic-Cap (age $\unit[8-34]{Myr}$) which is the association found to have hosted the potential encounter. The other three stars lie slightly below the model zero-age main sequence (ZAMS), thus were treated as ZAMS stars in T11. For HIP 79377 \citet{2009AA...501..941H} give an age of $\unit[200]{Myr}$. With a later spectral type, HIP 67655 and HIP 68783 are probably older. For those reasons, we do not consider HIP 67655, HIP 68783, HIP 74680 and HIP 79377 as former companion candidates for \rxjn{}. \\
At this stage, we do not exclude binary (and multiple) runaway systems as potential former companion candidates to the NS progenitor. The reason is that it is not excluded that a former binary (in rare cases also multiple) companion could have survived the SN explosion.\\
For the remaining eight stars -- HIP 40326, HIP 43158, HIP 57269, HIP 59803, HIP 76304, HIP 78078, HIP 78106 and HIP 84794 -- we explore the distribution of $d_{min}$. We adapt the theoretically expected curve (equation 1 in T10) to the first part of the $d_{min}$ distribution. \\
Three stars appear to have possibly been at the same place as \rxjn{} (i.e. $\mu=0$): HIP 43158, HIP 57269 and HIP 76304. In \autoref{tab:0720_3HIPs} we give the time and position of the SN and the properties \rxjn{} would currently have if it was born in the respective association with the runaway star being the former companion. In the last column of that table an estimation of the mass of the progenitor star is given derived from the age of the runaway star (see below) minus the time $\tau$ since the potential SN using evolutionary models from \citet{1980FCPh....5..287T}, \citet{1989A&A...210..155M} and \citet{1997PhDT........31K}. \\
In the case of HIP 43158 we find the radial velocity of \rxjn{} to be rather small. For that reason, we repeat the calculations adopting a uniform radial velocity distribution in the range of $\unit[-300\ldots+300]{km/s}$ (cf. \rxja{} and US, \autoref{subsec:1856parentassoc}). For HIP 43158, the results for both radial velocity distributions are given in \autoref{tab:0720_3HIPs}.\\
HIP 43158 is a single B0 giant star with a peculiar space velocity of $\approx\unit[53]{km/s}$ and an age of $\approx\unit[15]{Myr}$ (T11) which is consistent with an origin in Tr 10 (age $\unit[15-35]{Myr}$, see T10 and references therein). It has previously been suggested that HIP 43158 originates from the Vela region \citep{2008A&A...489..105S}. The position of the proposed SN (from uniform $v_r$, case $\#$: $l=\unit[258{.}9\ldots259{.}9]{deg}$, $b=\unit[1{.}8\ldots2{.}5]{deg}$) (that occurred $\approx\unit[0{.}9]{Myr}$ ago) lies in a region of enhanced $^{26}$Al emission \citep[][their figure 2, $^{26}$Al lifetime $\approx1\,\mathrm{Myr}$]{2010A&A...522A..51D}. Since $^{26}$Al is produced by Wolf-Rayet winds as well as SN explosions \citep[for a review see][]{1996PhR...267....1P}, we expect to find $^{26}$Al emission at the position of a recent SN.\\
HIP 57269 (spectral type K1V) is a double star in a triple system \citep{2003AN....324..516K} with a peculiar space velocity of $\approx\unit[21]{km/s}$ and an age of $\approx\unit[20]{Myr}$ (T11) that is still in its pre-main sequence phase \citep{2003AN....324..516K}. Its age is consistent with that of Tuc-Hor ($\unit[10-40]{Myr}$, see T10 and references therein), the potential host association of the SN. Comparing the position of the proposed SN ($l=\unit[281{.}3\ldots291{.}1]{deg}$, $b=\unit[15{.}3\ldots20{.}6]{deg}$) (that occurred $\approx\unit[0{.}3]{Myr}$ ago) with the map of $^{26}$Al $\gamma$-rays \citep[][]{2010A&A...522A..51D}, there seems to be only little $^{26}$Al emission in this area. Also, it might be doubtful whether a triple system could have survived a SN in a multiple system.\\
The suspected spectroscopic binary HIP 76304 (\citealt{2007AA...464..377F}; spectral type G2V) is an X-ray source and is listed as T Tauri star in the catalogue of T Tauri stars in Scorpius-Centaurus (Sco-Cen) by \citet{2000AA...356..541K} but with unknown classification. It has a peculiar space velocity of $\approx\unit[45]{km/s}$. \citet{2009AA...501..941H} give an age of $3{.}5$ to $\unit[5{.}9]{Gyr}$ which would be far too old to be associated with Sco-Cen or another young nearby association. Since it is an X-ray source, a pre-main sequence star with an age of $\approx\unit[9]{Myr}$ (T11) seems more plausible. Then its age would be consistent with that of $\beta$ Pic-Cap ($\unit[8-34]{Myr}$, see T10 and references therein), the potential host association of the SN. At the position of the supposed SN ($l=\unit[336{.}1\ldots341{.}4]{deg}$, $b=\unit[31{.}5\ldots33{.}2]{deg}$) (that occurred $\approx\unit[0{.}6]{Myr}$ ago), an enhancement of $^{26}$Al emission \citep{2010A&A...522A..51D} seems to be present.\\
The rotational velocities ($v\sin i$) of these three stars are rather small; $v\sin i=\unit[96\pm15]{km/s}$, $\unit[20]{km/s}$ and $\unit[5]{km/s}$ for HIP 43158, HIP 57269 and HIP 76304, respectively \citep{1996ApJ...463..737P,1993AA...273..194R,2009AA...501..941H}. If one of them originated from a supernova in a multiple system, this may indicate a small inclination angle $i$.\\
Considering that in the case of HIP 57269 only little $^{26}$Al emission is visible although the potential SN should have occurred very close to the Sun ($\approx\unit[27]{pc}$) and only $\approx\unit[0{.}3]{Myr}$ ago, this scenario seems less likely than for the other two candidate stars. Looking at the proposed parallax \rxjn{} would currently have in each case, also HIP 76304 is a less good candidate since the parallax would be much larger ($\unit[5{.}6-6{.}8]{mas}$) than the measured value ($\unit[3{.}6\pm1{.}6]{mas}$, \citealt{ThomasPhD}).\\
We suggest that \rxjn{} may have been born in a SN $\unit[0{.}85\pm0{.}15]{Myr}$ ago as a former member of Tr 10 with HIP 43158 being the possible former companion. It has been suggested that BSS runaway stars should be blue stragglers due to mass transfer during binary evolution, i.e. they appear younger, hence bluer, than their parent association (see also \citealt{2001A&A...365...49H} for other examples). In \autoref{fig:Tr10HRD}, we show the positions of HIP 43158 and Tr 10 member stars from \citet{1999AJ....117..354D} along with isochrones for $\unit[15]{Myr}$ and $\unit[35]{Myr}$ taken from \citet{2008A&A...482..883M}. If Tr 10 is only as young as $\unit[15]{Myr}$, HIP 43158 is not a blue straggler; however, Tr 10 can be as old as $\unit[35]{Myr}$ (T10 and references therein). In this case HIP 43158 would be a blue straggler.\\
The mass of the progenitor star would have been $\unit[13-14]{M_\odot}$ (for an age of Tr 10 of $\unit[15]{Myr}$) to $\unit[7-9]{M_\odot}$ (for an age of Tr 10 of $\unit[35]{Myr}$) corresponding to a spectral type of B1 to B2/3 on the main-sequence. This is consistent with the progenitor star of \rxjn{} having an earlier spectral type than the earliest current member of Tr 10 \citep[four B3 stars,][]{1999AJ....117..354D}. Tr 10 has been previously suggested to host the birth place of \rxjn{} by \citet{2003A&A...408..323M} who considered the general direction of the NS's motion and \citet{2007ApJ...660.1428K} who investigated the probability of close approaches of the NS to any of the OB associations given in \citet{1999AJ....117..354D}. They varied the parallax within $\unit[2{.}8\pm0{.}9]{mas}$ and the radial velocity in the range $v_r = \pm0{.}935v_t$ ($0{.}935$ corresponds to $1\sigma$ in $v_r$ for random orientation, $v_t$ is the transverse velocity) and found a separation between the NS and the centre of Tr 10 of $\unit[17]{pc}$ $\unit[0{.}7]{Myr}$ ago for $v_r=\unit[-20\ldots+50]{km/s}$, not inconsistent with our more complete calculations.\\
\begin{table*}
\caption{Potential former companion stars of \rxjn{} and the respective parent association/cluster. For details on the determination of the predicted NS parameters given in Columns 4-8 (radial velocity $v_r$, proper motion $\mu_\alpha^*$ and $\mu_\delta$, space velocity $v_{sp}$, parallax $\pi$) see additional online supporting information. Columns 9-11 indicate the distance to the Sun $d_\odot$ and equatorial coordinates (J2000.0, as seen from the Earth at present) of the potential SN (at the time of the SN). The last Column gives an estimation of the mass of the progenitor star derived from the progenitor's life-time (age of the runaway star, see text, minus the time since the potential SN) using evolutionary models from \citet{1980FCPh....5..287T}, \citet{1989A&A...210..155M} and \citet{1997PhDT........31K}.\newline
For HIP 43158 the results using a $v_r$ distribution derived from the one of pulsar spatial velocities \citep{2005MNRAS.360..974H} (superscript $*$) as well as a uniform distribution in the range $v_r=\unit[-300\ldots+300]{km/s}$ (superscript $\#$) are shown (see text).}\label{tab:0720_3HIPs}
\begin{tabular}{c c >{$}c<{$} >{$}r<{$} o{4.2} o{4.2} >{$}r<{$} >{$}r<{$} >{$}c<{$} >{$}r<{$} >{$}r<{$} >{$}c<{$}}
\toprule
HIP & Assoc./cl. & \tau & \multicolumn{1}{c}{$v_r$}			& \multicolumn{1}{c}{$\mu_{\alpha}^*$} & \multicolumn{1}{c}{$\mu_{\delta}$} & \multicolumn{1}{c}{$v_{sp}$} &	\multicolumn{1}{c}{$\pi$}						&	d_{\odot} 	& \multicolumn{1}{c}{$\alpha$}	& \multicolumn{1}{c}{$\delta$} & M_{prog}\\ 
		&			& \mathrm{[Myr]}	& \multicolumn{1}{c}{[km/s]} & \multicolumn{1}{c}{[mas/yr]} & \multicolumn{1}{c}{[mas/yr]} & \multicolumn{1}{c}{[km/s]} & \multicolumn{1}{c}{[mas]} & \multicolumn{1}{c}{[pc]} & \multicolumn{1}{c}{[$^\circ$]} & \multicolumn{1}{c}{[$^\circ$]} & \mathrm{[M_\odot]}\\\midrule
43158$^*$	&	Tr 10	& 0{.}82\ldots1{.}18	&	-100^{+25}_{-5}	&	-92{.}6+1{.}3	&	55{.}9+1{.}7	&	160^{+10}_{-10}	&	3{.}6^{+0{.}8}_{-0{.}2}	&	370^{+4}_{-18}	&	130{.}2^{+0{.}9}_{-0{.}6}	&	-38{.}7^{+0{.}3}_{-0{.}3} & \approx13-15\\
43158$^\#$	&	Tr 10	& 0{.}70\ldots1{.}00	&	-76^{+34}_{-17}	&	-92{.}8+1{.}4	&	55{.}5+1{.}6	&	163^{+3}_{-8}	&	3{.}5^{+0{.}3}_{-0{.}3}	&	375^{+4}_{-16}	&	131{.}1^{+0{.}2}_{-1{.}1}	&	-38{.}8^{+0{.}4}_{-0{.}2} & \approx13-15\\
57269	&	Tuc-Hor		& 0{.}20\ldots0{.}33	&	511^{+143}_{-109}	&	-92{.}7+1{.}5	&	55{.}4+1{.}8	&	592^{+103}_{-150}	&	6{.}3^{+0{.}4}_{-1{.}0}	&	27^{+3}_{-3}	&	171{.}1^{+8{.}4}_{-6{.}4}	&	-42{.}4^{+0{.}6}_{-1{.}2} & \approx10-12\\
76304	&	$\beta$ Pic-Cap	& 0{.}52\ldots0{.}70	&	309^{+27}_{-69}	&	-92{.}8+1{.}4	&	55{.}2+1{.}6	&	338^{+29}_{-81}	&	5{.}8^{+1{.}0}_{-0{.}2}	&	52^{+4}_{-2}	&	226{.}5^{+1{.}6}_{-1{.}9}	&	-20{.}9^{+1{.}6}_{-1{.}4} & \approx19-33\\
\bottomrule
\end{tabular}
\end{table*}

Since we cannot pre-constrain the radial velocity of \rxjn{}, it is difficult to examine those 842 runaway star candidates in T11 without radial velocity measurements due to the large uncertainties (assuming $v_r=\unit[-500\ldots500]{km/s}$). For already 304 of the 842 stars it is possible to find a past position as close as $\unit[10]{pc}$ to \rxjn{} after $10^4$ Monte Carlo runs. After further three million runs, we found for 113 of those 304 stars a smallest separation to \rxjn{} that was less than one parsec (for justification of the limits see additional online supporting information). Excluding stars that would need a peculiar space velocity larger than $\unit[180]{km/s}$ (that is $\approx6\sigma$ above the maximum of the distribution of runaway star velocities, T11), 34 candidates are left. For all of them we can find a set of input parameters ($\pi$, $\mu^*_\alpha$, $v_r$) for that we can find close encounters ($d_{min}\leq\unit[10]{pc}$) with \rxjn{} within one or more of the nine potential birth places listed in \autoref{tab:assoc0720_2}. Here, constraining the runaway star radial velocity through observations is necessary.\\
Most of the 34 stars are late-type stars that are situated in the HRD on or near the ZAMS. For that reason, we doubt that they are indeed sufficiently young runaway stars for \rxjn{} ($\lesssim$ few Myr).\footnote{In T11, they were treated as ZAMS stars and contribute to the 20 per cent contamination that are no true runaway stars but fast-moving members of the low-velocity group.} 
%______________________________________________________________

\section{Summary and conclusions}\label{sec:summary}

We analysed the origin of two members of the M7, \rxja{} and \rxjn{} using most recent parallax measurements. \\
Under the assumption that most NSs are born in and ejected from their parent association or cluster (see \autoref{sec:intro}), we confirm that \rxja{} most probably originated from the US association as suggested previously (\citealt{2002ApJ...576L.145W}, T10). We find that the radial velocity of \rxja{} is $\unit[6^{+19}_{-20}]{km/s}$ implying an inclination angle to the line-of-sight of $\unit[88\pm6]{deg}$ that is consistent with the inclination of the bow shock that \rxja{} creates in the ISM owing to its motion \citep{2001A&A...380..221V}. This consistency implies that the current distance of \rxja{} is $\approx\unit[120]{pc}$ as obtained by \citet{2002ApJ...576L.145W} and \citet{2010ApJ...724..669W} rather than $\unit[160-180]{pc}$ \citep{2003conf...Kaplan,2007Ap&SS.308..191V}. This smaller distance is also in good agreement with \rxja{} being in front of the Corona Australis star forming region ($\approx\unit[130]{pc}$, \citealt{2008hsf2.book..735N}); if it was behind, optical detection would probably have failed. Moreover, this is in accordance with the absence of OI from the ISM in the Chandra HRC/LETG spectra \citep{2001A&A...379L..35B,2003A&A...399.1109B}. The space velocity of \rxja{} would then be $\approx\unit[195]{km/s}$. The derived kinematic age of \rxja{} of $\unit[0{.}46\pm0{.}05]{Myr}$ agrees well with theoretical cooling models (cf. T10, Figure 12 therein). \\
The mass of the progenitor star would have been $\approx\unit[40-60]{M_\odot}$. Considering that the NS may have originated from a binary system that got disrupted in an asymmetric SN explosion, we searched for a possible former companion that should then be a runaway star. No former companion candidate was identified. A reason could be that, if there was a former companion, either, it may not be included in the runaway star catalogue of T11 either because it was no Hipparcos source or its velocity vector is not significantly different from those of its neighbouring stars (note, that T11 also investigated the direction of motion of young stars compared to neighbouring stars, regardless whether the absolute velocity is large or small); or a former companion might also have experienced a SN after ejection (then, the mass ratio of the binary would have been close to unity). Alternatively, the progenitor star of \rxja{} could have been a single star and the NS gained its high velocity in an asymmetric SN explosion. However, single stars of masses $\gtrsim\unit[20-25]{M_\odot}$ are generally believed to produce black holes rather than NSs. \citet{2001ApJ...560L..83M}, \citet{2002A&A...390..299B} and \citet{2009SSRv..143..437F} suggested that about 20 nearby (in Sco-Cen) SNe created and re-heated the Local Bubble. \rxja{} was probably formed in one of them about half a million years ago in US. For that reason and since the small radial velocity of \rxja{} that we found is strongly supported by the bow shock measurements of \citet{2001A&A...380..221V}, we argue that \rxja{} is not a remnant of a runaway star that had already left its parent cluster when it experienced a SN.\\

For \rxjn{} there is no unique result on the parent association due to the large parallax uncertainties (as long as runaway stars are not concerned). It could have been formed either in one of the young local associations \citep{2008A&A...480..735F} with TWA being the best candidate birth place (see also T10) or in a more distant association/cluster such as Tr 10 or Col 140. In the first case, the distance of the SN to the Sun is very small ($\approx\unit[30-50]{pc}$) and it is expected to find SN-produced radionuclides on Earth \citep{1996ApJ...470.1227E}. A small but insignificant signal of $^{60}$Fe was found for $\unit[0-1]{Myr}$ in the past \citep{2004PhRvL..93q1103K}. \\
The identification of another indicator is thus needed. We find three runaway stars for which not only a positional coincidence with \rxjn{} is possible but also an enhancement of $^{26}$Al emission was measured at the location of the potential SN \citep{2010A&A...522A..51D}. Regarding the current distance of \rxjn{} derived for each case, we suggest that \rxjn{} may have been born $\unit[0{.}85\pm0{.}15]{Myr}$ ago as a former member of Tr 10 with HIP 43158 being the possible former companion. The mass of the progenitor star would have been $\approx\unit[8-14]{M_\odot}$. The current distance of \rxjn{} would then be $\approx\unit[285]{pc}$, in good agreement with estimates from $n_H$ measurements \citep[$250\pm25\,\mathrm{pc}$,][]{2007Ap&SS.308..171P} and the most recent HST parallax (distance $\unit[280^{+210}_{-85}]{pc}$, \citealt{ThomasPhD}). The present radial velocity of \rxjn{} would be $\unit[-76^{+34}_{-17}]{km/s}$ leading to a space velocity of $\approx\unit[160]{km/s}$.\\
However, we emphasise that further observation of the former companion candidate is needed to confirm its status. There are also runaway stars without radial velocity measurements that are potential former companion candidates for \rxjn{}. For them, a measurement of the radial velocity is crucial.\\
\rxjn{} is probably older than \rxja{}, thus should be cooler than \rxja{}. However, the effective temperature of \rxjn{} is $kT\approx\unit[85-95]{eV}$ \citep{2009A&A...498..811H} while for \rxja{} it is $kT\approx\unit[60]{eV}$ \citep{2003A&A...399.1109B}. Hence, \rxja{} is cooler although it is probably younger. This might be due to different birth parameters or different cooling. Moreover, effective temperatures may be influenced by hot spots, hence do not reflect the real surface temperature. For \rxjn{}, this is indeed the case \citep{2006A&A...451L..17H,2009A&A...498..811H} and it appears hotter.

\section*{Acknowledgments}

We thank the referee, Vasilii Gvaramadze, for comments that helped to improve the manuscript.
NT acknowledges financial support from Carl-Zeiss-Stiftung.
MMH and TE acknowledge support from DFG in the SFB/TR-7 Gravitational
Wave Astronomy. We thank Fred Walter for reading the manuscript and useful comments and suggestions on that as well as for discussion on \rxja{}. We thank Roland Diehl for useful discussion on $\gamma$ observations. We would also like to thank Frank Haberl and Valeri Hambaryan for suggestions that helped in improving the manuscript as well as Amalya Poghosyan for providing an updated catalogue of SN remnants (not yet public). We thank Lukasz Bukowiecki for discussions on cluster data and Thomas Preibisch on discussions on Upper Scorpius in particular.\\
Our work has made use of the Simbad and VizieR services (http://cds.u-strasbg.fr). 
\bibliographystyle{mn}
\bibliography{bib_1856Paper}

\label{lastpage}

\end{document}